\documentclass[aps,prd,twocolumn,showpacs,groupedaddress,preprintnumbers,superscriptaddress,amsmath,amssymb,floatfix,longbibliography]{revtex4-1} 
 
\def\nn{\nonumber}

\newcommand{\Npup}[0]{ N_{\rm upper} }

\newcommand{\Avrg}[1]{\left\langle #1 \right\rangle}

\usepackage{enumerate}
\usepackage{multirow}
\usepackage{float}

\usepackage{xcolor}

\usepackage{color}			
\usepackage{graphicx}
\usepackage[export]{adjustbox}
\DeclareGraphicsExtensions{.png,.jpg}
\graphicspath{}	
\usepackage[section]{placeins} 

\usepackage{physics}
\usepackage{braket}
\usepackage{xparse}	

\usepackage{slashed}
\usepackage{indentfirst} 
\usepackage{enumitem}

\usepackage[colorlinks=true, filecolor=blue, citecolor=blue,urlcolor=blue]{hyperref}
\usepackage{url}
\usepackage{orcidlink}

\date{\today}

\begin{document}

\abovedisplayskip = 4pt
\belowdisplayskip = 4pt
\abovedisplayshortskip = 4pt
\belowdisplayshortskip= 4pt

\title{Prospects of boosted  magnetic dipole inelastic fermion dark matter at ILC-BDX} 

\author{I.~V.~Voronchikhin \orcidlink{0000-0003-3037-636X}}
\email[\textbf{e-mail}: ]{i.v.voronchikhin@gmail.com}
\affiliation{Institute for Nuclear Research, 117312 Moscow, Russia}
\affiliation{ Tomsk Polytechnic University, 634050 Tomsk, Russia}

\author{D.~V.~Kirpichnikov \orcidlink{0000-0002-7177-077X}}
\email[\textbf{e-mail}: ]{dmbrick@gmail.com}
\affiliation{Institute for Nuclear Research, 117312 Moscow, Russia}

\begin{abstract}
In this work, we investigate the projected sensitivity of the Beam-Dump eXperiment at the International Linear Collider (ILC-BDX) to inelastic fermionic dark matter coupled to the Standard Model photon through an off-diagonal magnetic dipole operator.
We compute the production rate of dark matter states in the bremsstrahlung like  process \mbox{$e^- N \to e^- N \gamma^* (\to \chi_{1} \bar{\chi}_0)$}, induced by the scattering of high-energy electrons on target nuclei. 
The resulting boosted dark matter fluxes are then propagated to the detector, where the signal events arise from scattering off detector electrons or monophoton decay of heavier state. 
The projected exclusion limits are derived using the expected numbers of electrons on target (this implies a typical rate of \mbox{$4.0~\times~10^{21}/\mbox{year}$}) corresponding to 1 year and 10 years of data taking. 
To characterize the impact of inelasticity, we consider two benchmark relative mass splittings, \mbox{$\Delta=0.05$} and \mbox{$\Delta=0.001$}, motivated by thermal dark matter scenarios.
Our results show that ILC-BDX can probe inelastic magnetic-dipole dark matter over a phenomenologically viable region of parameter space.
\end{abstract}

\maketitle

\section{Introduction}

Observational evidence from cosmology and astrophysics strongly supports the existence of dark matter (DM), providing one of the clearest indications of physics beyond the Standard Model (SM)~\cite{Planck:2018vyg}. However, the specific particle nature of DM and the structure of its interactions with the visible sector remain unknown~\cite{PBC:2025sny}.

Thermalization of DM with the visible sector relates the present-day cosmological abundance to microscopic interactions in the early Universe~\cite{Gondolo:1990dk}. Considerable attention has focused on light thermal DM, for which a freeze-out mechanism is required to deplete the relic abundance to the observed value. Such a mechanism can be realized through a hidden mediator of various possible spins~\cite{Voronchikhin:2026ffj,Wang:2025xoq,Izaguirre:2015yja,Betancur:2026fyh,Kang:2020huh,Voronchikhin:2024ygo}.
However, instead of introducing a new light sub-GeV portal interaction, one can consider couplings of DM directly to the ordinary photon~\cite{Dienes:2023uve,Jodlowski:2023ohn,Kling:2022ykt,Chu:2018qrm}. 

\begin{figure*}
\centering
\includegraphics[width=0.9\textwidth]{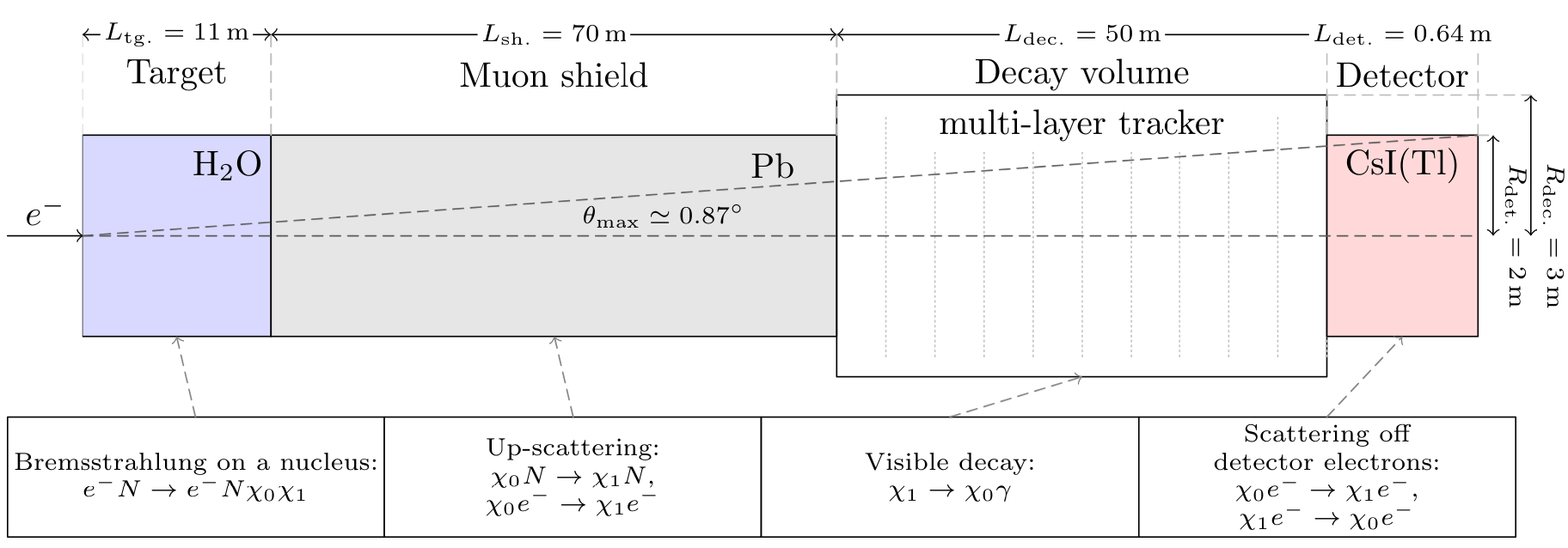}
\caption{ The ILC-BDX detector setup~\cite{Asai:2021ehn,Asai:2023dzs}. 
\label{fig:detSeup}}
\end{figure*}

Magnetic dipole dark matter is one of the Standard Model extensions in which the hidden sector is coupled to the visible sector through the electromagnetic photon within an effective field theory framework~\cite{Chu:2018qrm}. 
Moreover, magnetic dipole dark matter possesses the largest unconstrained region of parameter space among the different electromagnetic form-factor components~\cite{Kling:2022ykt,Jodlowski:2023ohn}. 
However, the case of elastic light magnetic dipole dark matter is strongly constrained by the observed spectrum of cosmic microwave background anisotropies~\cite{Chu:2018qrm}. 
Introducing a mass splitting~\cite{Tucker-Smith:2001myb} between hidden-sector particles can lead to suppressing indirect and direct-detection
signatures, where the electromagnetic operator coupling two nearly degenerate dark-sector states~\cite{Dienes:2023uve}.

Specifically, we consider an inelastic dark matter (iDM) scenario with 
two 4-component Majorana 
  states, $\chi_0$ and $\chi_1$,  described by the Lagrangian~\cite{Batell:2021ooj,Dreiner:2008tw}
\begin{equation}
\mathcal{L}^{\rm DM}_{\rm kin.} \supset  \sum_{i = 0,1} \left[ \frac{1}{2} \overline \chi_i \, i \gamma^\mu  \,\partial_\mu \chi_i
-\frac{1}{2} m_{\chi_i}\, \overline \chi_i \chi_i \right]\, ,
\label{KinTerMajoranaDM}
\end{equation}
where $m_{\chi_i}$ denotes the physical fermion masses, satisfying $m_{\chi_0}~<~m_{\chi_1}$. We introduce the relative mass splitting:
\begin{equation}
\Delta = (m_{\chi_{1}} - m_{\chi_{0}})/m_{\chi_{0}} \geq  0.
\label{SplittinDefinition}
\end{equation}

These states are coupled to the Standard Model photon via an off-diagonal dimension-five magnetic dipole operator.
The effective Lagrangian includes the 
term~\cite{Chang:2010en,Jodlowski:2023ohn,Dienes:2023uve}
\begin{equation}
\label{LagrangianMDM}
\mathcal{L} \supset \frac{1}{\Lambda_{\rm M}}\, i \bar{\chi}_1 \sigma^{\mu \nu} \chi_0 F_{\mu \nu},
\end{equation}
where the scale $\Lambda_{\rm M}$ characterizes the strength of 
this non-renormalizable interaction from new physics, 
$F_{\mu \nu}$ denotes the electromagnetic field strength 
tensor, and $\sigma^{\mu\nu}=i [\gamma^\mu,\gamma^\nu]/2$
represents the commutator of the Dirac gamma matrices.

For DM masses below $\mathcal{O}(1)~\mathrm{GeV}$, accelerator-based 
searches combine the benefits of sufficiently high beam energies and 
relatively large primary-beam intensities (for a recent review, see 
e.~g.~Refs.~\cite{Izaguirre:2017bqb,Mongillo:2023hbs,PBC:2025sny,Feng:2022inv,Gninenko:2026svu,NA64:2021acr,Voronchikhin:2025eqm} and references therein).
The introduction of a small mass splitting also enriches the phenomenology of the model, giving rise in particular to up-scattering, down-scattering, and decays of the heavier state~\cite{Jodlowski:2023ohn,Jodlowski:2023yne}. Thus, the model of inelastic light magnetic dipole dark matter leads to specific experimental signatures that can be probed in fixed-target experiments.

The ILC-BDX is a proposed beam-
dump facility at the future energy-frontier linear $e^+e^-$-collider planned to be 
constructed in the Kitakami area, Tohoku, Japan.
The ILC-BDX experiment is expected to have strong sensitivity to a light dark 
sector~\cite{Zarnecki:2020ics,Asai:2021ehn,Nojiri:2022xqn,Asai:2023dzs}; however, its 
sensitivity to inelastic magnetic dipole dark matter~\cite{Chang:2010en,Jodlowski:2023ohn,Dienes:2023uve} has not yet been investigated in 
detail.  
In this paper, we therefore focus on expected reach of the 
inelastic  magnetic dipole dark matter model through its (i) scattering off 
electrons; (ii) monophoton decay in  the proposed ILC-BDX  detector.

This paper is organized as follows. 
In Sec.~\ref{sec:ExperimentSetup} we outline the  benchmark ILC-BDX experiment. 
In Sec.~\ref{sec:RelAbCalcEq} we briefly summarize the main cosmological and astrophysical implications of the model under consideration.
In Sec.~\ref{sec:SignalInILC} we discuss the rate of 
iDM pair production at ILC-BDX associated with bremsstrahlung-like emission. 
In Sec.~\ref{sec:ExpectedReach}, we derive the projected exclusion limits 
of the ILC-BDX fixed-target facility in the  iDM  parameter space. 
We conclude in~Sec.~\ref{sec:Conclusion}.

\section{The benchmark Experiment
\label{sec:ExperimentSetup}}

In this section, we  summarize experimental parameters of considered fixed-target facility
(see Fig.~\ref{fig:detSeup}).

We assume that the ILC-BDX experiment 
operates with a primary beam of energy $250~\mathrm{GeV}$~\cite{Fujii:2017vwa,Zarnecki:2020ics} or~$125~\mathrm{GeV}$~\cite{Asai:2021ehn}  in the laboratory frame and an 
intensity corresponding to $4.0~\times~10^{21}$ electrons accumulated on target 
per year~\cite{Asai:2021ehn}. Downstream of an $11$ m-long water beam dump,~$L_{\rm tg.}~=~11~\mbox{m}$, there is a lead muon shield of length $L_{\rm sh.}~=~70~\mbox{m}$, followed by a decay volume of length $L_{\rm dec.}~=~50~\mbox{m}$ instrumented with a multilayer tracking system~\cite{Asai:2023dzs}. The tracking system in the decay volume can be used for 
searches for visible dark matter decays. 
The decay volume is followed by a cylindrical 
CsI(Tl) electromagnetic calorimeter with radius of $2~\mathrm{m}$ and length of $0.64~\mathrm{m}$, 
characterized by an electron number density $n_{\rm det}^{e}~=~1.1~\times~10^{24}~\mbox{cm}^{-3}$.

The sufficiently large shielding volume implies that the relevant background events are 
primarily induced by the neutrino flux produced along the beamline and in the muon shield. 
The irreducible background for this setup arises from elastic scattering of beam-originated neutrinos off electrons in the calorimeter, which can mimic a signal-like electromagnetic shower. 
For a threshold energy of $E_{\rm th}=1~\mathrm{GeV}$, the 
contribution of the irreducible background is estimated to be at the level of $\sim~\mathcal{O}(1)$ event per year. 
In particular, the corresponding upper limits on the number of signal events are $\Npup~=~3.8$ and $\Npup~=~7.3$ at $95\%$ CL for statistics of 1 and 10 years, respectively.   
The background from cosmic neutrinos and muons is expected to be negligible~\cite{Asai:2023dzs}. 

The dominant reducible background channels are neutrino-nucleon interactions and neutral-pion production induced by beam-originated neutrinos in the detector, which can mimic electron-recoil signals. 
However, reducible background can potentially be suppressed through an analysis of the electromagnetic-shower profile~\cite{Asai:2021ehn}.
To account for all uncertainties in the final detector design and analysis 
performance, we illustrate the exclusion limits with  
100 and 1000 expected signal events~\cite{Asai:2023dzs}. 

As was pointed out above, the high beam energy and intensity of this setup can provide a viable sensitivity to dark matter 
models. 
We focus on a scenario of inelastic dipole dark matter, in which a flux of $\chi_i$ particles is produced 
when the primary electrons impinge on the target. 
The resulting boosted dark matter~\cite{Kim:2016zjx,Agashe:2014yua} can 
reach the detector and be probed through scattering in the detector 
material~\cite{Batell:2014mga} or the decay of the heavy state in the fiducial 
volume Ref.~\cite{Jodlowski:2023ohn,Jodlowski:2023yne}.

\section{Relic abundance of inelastic magnetic dipole dark matter
\label{sec:RelAbCalcEq}}

In this section, we discuss the relic abundance of inelastic magnetic dipole dark matter in an analytic approximation, where the calculation of the relic curves is reduced to solving the Boltzmann equation for the total number density of hidden-sector particles~\cite{Griest:1990kh,Gondolo:1990dk}.
We also summarize the main astrophysical constraints on the parameter space of the model under consideration~\cite{Chu:2018qrm}.

\begin{figure*}[tbh!]
\centering
\includegraphics[width=0.7\textwidth]{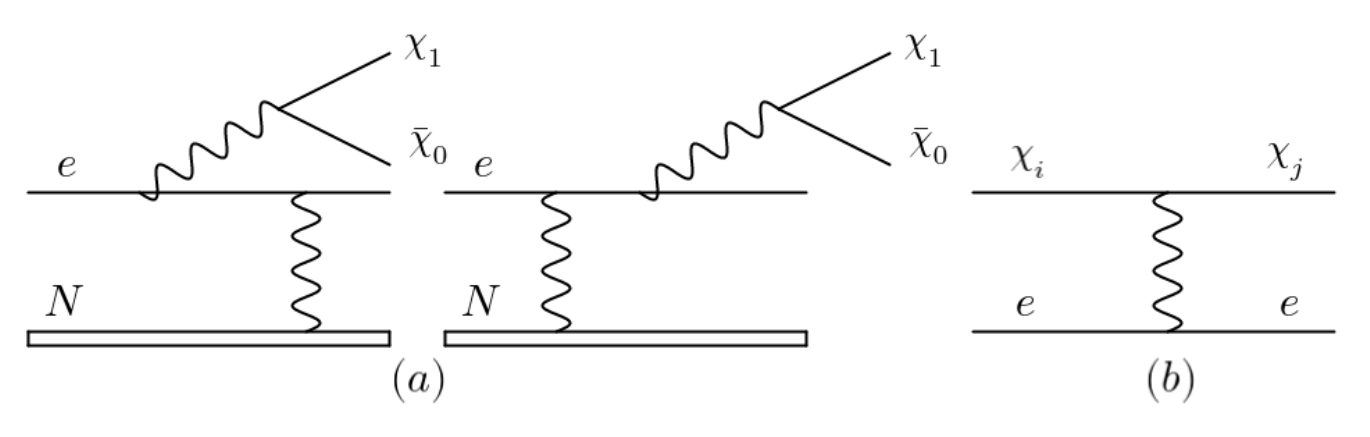}
\caption{ Feynman diagrams illustrating the iDM processes of interest: (a) bremsstrahlung-like production of iDM  via an off-shell photon emission $e N \to e N \gamma^*(\to \chi_1 \bar{\chi}_0)$; (b) detection reaction of boosted iDM via its scattering off electrons $\chi_i e\to \chi_j e$. 
\label{FeynamProduction}}
\end{figure*}

\subsection{The relic abundance curves}

Assuming kinetic equilibrium between the hidden and visible sectors~\cite{Binder:2021bmg} and chemical equilibrium within the hidden sector~\cite{Garny:2017rxs}, the Boltzmann equation for the total number density take the form~\cite{Kolb:1990vq,Dienes:2023uve,Jodlowski:2023ohn,Voronchikhin:2025eqm}:
\begin{align}
    \dot{n} 
=  
    & -   3 H n
 -
        \Avrg{\sigma_{\rm eff} v}
        \left( n^2 - n_{\rm eq}^2\right),
\end{align}
where $n~=~\sum_{i=0,1} n_i$~is a total particle density of dark matter.
The effective thermally averaged cross section is~\cite{Edsjo:1997bg}:
$$
    \Avrg{\sigma_{\rm eff} v}
=
    \frac{m_{\rm \chi_0}}{8 \pi^4 x}
    \frac{1}{n_{\rm eq}^2}
    \int\limits_{(m_0 + m_1)^2}^{\infty}
    \sigma_{\rm eff}(s)
    K_1\left( \frac{\sqrt{s}}{m_{\chi_0}} x \right)
    d s,
$$
where $K_i(s)$ is the modified Bessel function of the second kind, $x = m_{\chi_1}/T$ and is denoted:
$$
    \sigma_{\rm eff} 
=  
    \sum\limits_{f}
    \sum\limits_{i,j=0,1}
    g_i g_j  \frac{\lambda(s, m_{\chi_i}^2, m_{\chi_j}^2)}{4\sqrt{s}}  \sigma_{ij\to f}(s),
$$
$\sum_{f}$ is the sum over corresponding final states and $\lambda(s, m_{\chi_i}^2,m_{\chi_j}^2)$ is a Kallen function.
The above expression includes the coannihilation process~$\chi_0 \chi_1~\to~{ll}$, where the total cross section are given in Ref.~\cite{Dienes:2023uve}. 
We neglect the dark matter annihilation processes, since this contribution are suppressed by additional powers of the interaction coupling~\cite{Jodlowski:2023ohn}.
We also take into account coannihilation into hadrons using the expression~\cite{Krnjaic:2025noj}:
$$
\sigma_{\chi_0 \chi_1 \to {\rm hadrons}} = R(\sqrt{s})\sigma_{\chi_0 \chi_1 \to \mu \mu},
$$
where R-ratio,~$R(\sqrt{s})$, can be found in the Ref.~\cite{ParticleDataGroup:2022pth}.

The current value of the cold DM relic abundance obtained from 
the Planck 2018 combined analysis is~\cite{Planck:2018vyg,Husdal:2016haj}:
\[
    \Omega_c h^2 = 0.1200 \pm 0.0012.
\] 
The expression of relic density reads~\cite{Kolb:1990vq}:
$$
    \Omega_{c} 
\simeq 
    \frac{ m_{\chi_1} s_0}{3 M_{\rm Pl}^2 H_0^2}
    \frac{H(m_{\chi_1})}{\; s(T = m_{\chi_1})}
    \left(\, \int\limits_{x_f}^{\infty} \frac{\Avrg{\sigma_{\rm eff} v}}{x^2} dx \right)^{-1} \!\!, 
$$
where $g_{*\varepsilon}(T)$,~$g_{*s}(T)$ is the energy and entropy effective degrees of freedom~\cite{Husdal:2016haj}, respectively, $s_0 = s(T_0)$ is the current total entropy density and $T_0$ is the current temperature of the Universe. 

We note that our calculations based on the analytic approximation for the mass splittings $\Delta=0.001$ and $\Delta = 0.05$ agree with the numerical solution of the system of Boltzmann equations presented in Ref.~\cite{Jodlowski:2023ohn,Dienes:2023uve}.

 \begin{figure}[t!]
\centering
\includegraphics[width=0.25\textwidth]{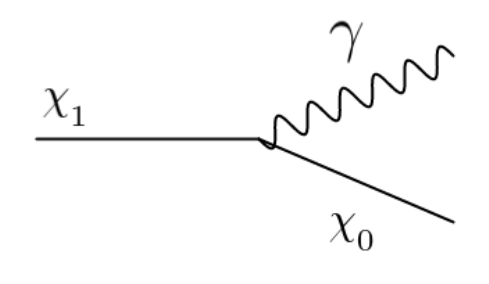}
\caption{ Feynman diagram describing two-body radiative decay of excited iDM state, $\chi_1$, into SM photon, $\gamma$, and lightest mode $\chi_0$. 
\label{FeynamDecay}}
\end{figure}

\subsection{Astrophysical constraint}

The anisotropies of the cosmic microwave background (CMB) are sensitive to the energy injected into the intergalactic medium by dark matter annihilation near the epoch of recombination.
Therefore, this leads to constraint on the thermal averaged cross section~\cite{Gondolo:1990dk,Voronchikhin:2025eqm} at the level of~\cite{Galli:2013dna,Planck:2018vyg}:
\begin{align}
    \langle\sigma v\rangle_{\rm s-wave}^{\rm CMB}    
< 
    \left(\frac{3.2\times10^{-28}}{f_{\rm eff}}\right)
    \left(\frac{m_{\rm DM}}{{\rm GeV}}\right)
    {\rm cm^3/s}.
\end{align}
where $f_{\rm eff}~\simeq~0.5$ is the effective power fraction that is injected by the annihilation of DM and
deposited in the intergalactic medium~\cite{Slatyer:2009yq,Slatyer:2015jla}.
For magnetic dipole inelastic dark matter, the cross section of the t-channel annihilation process $\chi_0\chi_0 \to \gamma \gamma$ is~\cite{Jodlowski:2023ohn}:
\begin{align}
	\left\langle \sigma_{\chi_0 \chi_0\to\gamma\gamma} v \right\rangle
\simeq
    \frac{16}{\pi\Lambda_{\rm M}^4}
    \frac{m_{\chi_0}^4 m_{\chi_1}^2}{(m_{\chi_0}^2+m_{\chi_1}^2)^2}.
\end{align}
The annihilation cross section of the lightest dark matter particle is s-wave dominated, which lead to constraints as
\begin{align}
    \frac{1}{\Lambda_{\rm M}}
< &
    2.35\times 10^{-3}\;{\rm GeV}^{-1}
    \left(\frac{1~{\rm GeV}}{m_{\chi_0}}\right)^{1/4}
\nn \\ \; & \times
    \left(\frac{0.7}{f_{\rm eff}}\right)^{1/4}
    \left[
    \frac{\Delta^2+2\Delta+2}
    {2(1+\Delta)}
    \right]^{1/2}.
\end{align}
Thus, the strongest CMB constraints are obtained for masses around~$1~\mbox{GeV}$, at the level $1/\Lambda_{\rm M}~<~2.35~\cdot~10^{-3}$. This behavior arises because dark matter annihilation is suppressed as $\sigma_{\chi_0 \chi_0\to\gamma\gamma}~\sim~(1/\Lambda_{\rm M})^{4}$, while the relic abundance in the case of~$\Delta~<~\mathcal{O}(1/x_{f})$ is set by the more efficient coannihilation process~\cite{Jodlowski:2023ohn}.
It is also worth noting that the annihilation channel $\chi_0\chi_0\to\gamma\gamma$ can lead to a monochromatic photon signal at gamma-ray telescopes~\cite{Bartels:2017dpb}. 
In particular, for the mass range~\mbox{$10~{\rm MeV}-\mathcal{O}(1)~{\rm GeV}$}, the corresponding bounds are in the range: 
$$
2 \cdot 10^{-3}~\mbox{GeV}^{-1}   \gtrsim 1/\Lambda_{\rm M} \gtrsim 9 \cdot 10^{-4}~\mbox{GeV}^{-1}  ,
$$
for both mass splitting as $\Delta = 0.05$ and $\Delta = 0.001$.
Therefore, annihilation of the lighter state does not impose any new constraints on the corresponding region of parameter space~\cite{Dienes:2023uve}.

Next, the parameter space of elastic magnetic dipole dark matter in the sub-GeV mass range is excluded by the observed CMB spectrum bounds~\cite{Chu:2018qrm}.
We also note that coannihilation,~$\chi_0 \chi_1\to ll$, is s-wave process~\cite{Jodlowski:2023ohn} and can  lead to constraints from the observed CMB anisotropies. 
However, we expect, that such constraints are weak because of the suppressed abundance of the heavier dark-sector state for mass splitting~\mbox{$\Delta \simeq \mathcal{O}(10^{-3})-\mathcal{O}(10^{-2})$}~\cite{CarrilloGonzalez:2021lxm}. 
Since the $\chi_1$ number density is efficiently depleted after the chemical freeze-out in the hidden sector. 

It is also worth noting that the channel \mbox{$\chi_0 \chi_0 \to \chi_1 \chi_1$} is kinematically open only for velocities satisfying \mbox{$v^2>\mathcal{O}(\Delta)$} in the regime of small mass splittings, \mbox{$\Delta \ll 1$}.
In particular, the self-interaction for \mbox{$\Delta < \mathcal{O}(10^{-5})$} can lead to constraints at the level of \mbox{$\sigma/m_{\chi_0}<\mathcal{O}(1)~\mbox{cm}^2/\mbox{g}$}~\cite{Markevitch:2003at,Harvey:2015hha,Jee:2026qjg}.
Moreover, in the limit $\Delta \to 0$, the corresponding constraint arises at the level $1/\Lambda_{\rm M} \lesssim \mathcal{O}(10)~\mbox{GeV}^{-1}$~\cite{Chu:2018qrm} which  lies outside the regime of effective-field-theory description.
Thus, we assume that the constraints from dark matter self-interactions can be considered negligible.

For sub-GeV dark matter, constraints from xenon-based direct-detection experiments can arise from scattering off atomic electrons~\cite{Essig:2017kqs}.
In particular, direct-detection constraints based on up-scattering,~\mbox{$\chi_0 e\to \chi_1 e$}, depend weakly on the mass splitting for~$\Delta~<~\mathcal{O}(10^{-7})$~\cite{Chu:2018qrm,Voronchikhin:2026ffj}.
Moreover, in the limit $\Delta \to 0$ and for masses $m_{\chi_0} > 10~\mbox{MeV}$  direct detection lead to constraint at the level \mbox{$1/\Lambda_{\rm M} \lesssim \mathcal{O}(10^{-5})$}, that excludes the corresponding region of parameter space of elastic magnetic dipole dark matter~\cite{Chu:2018qrm}.
However, for the mass splittings~$\Delta~>~\mathcal{O}(10^{-4})$, these up-scattering constraints are kinematically suppressed.

Thus, over the mass range from a $10$~MeV to a few GeV and mass splitting~$\Delta~\gtrsim~\mathcal{O}(10^{-3})$,  the dominant constraints on the parameter space of the inelastic dipole dark matter model arise from accelerator-based experiments.

\section{Signatures of  magnetic dipole inelastic dark matter \label{sec:SignalInILC}}

In this section, we discuss signal events at ILC-BDX induced by scattering of dark matter particles off detector electrons and decay of the heavy state in the fiducial volume.
In particular, scattering of the primary high-energy electron beam off target nuclei can produce dark matter via an off-shell Standard Model photon, while the emitted hidden-sector fermions carry away part of the beam energy.

\subsection{Boosted inelastic dark matter}

The dark matter production process is $2\to 4$ reaction~\cite{Chu:2018qrm,Dienes:2023uve} depicted in Fig.~\ref{FeynamProduction}~(a):
\begin{equation}
e^- N \to e^- N \gamma^* \to  e^- N   \bar{\chi}_0 \chi_1.
\label{eNtoeNChi0Chi1}
\end{equation}

We use the CalcHEP~\cite{Belyaev:2012qa} software package to calculate the iDM production cross sections (for detailed discussion of iDM model implementation in CalcHEP, see e.~g.~Refs.~\cite{Gninenko:2026svu,Gninenko:2026mgn,Gninenko:2025aek,Arefyeva:2022eba}). Specifically, 
we denote by~$\sigma_{2\to4}^{\rm cut}$ the cross section for the $2\to4$ process integrated over the dark matter energy $E_{\chi_0}$ with the angular cut $\theta_{\chi}^{\rm max}\simeq 0.874^{\circ}$.
Similarly, we introduce the total cross section for the considered process by $\sigma_{2\to4}^{\rm tot}$, which implies that no angle cut is imposed.
The corresponding dark-matter production cross sections, $\sigma_{2\to4}^{\rm cut}$ and $\sigma_{2\to4}^{\rm tot}$, are presented in Tabs.~\ref{tab:TCS2to4Ebeam125GeV} and~\ref{tab:TCS2to4Ebeam250GeV} for primary beam energies of 125 and 250~GeV, respectively.
We also provide their ratio~$\sigma_{2\to4}^{\rm cut}/\sigma_{2\to4}^{\rm tot}$ in these tables, which quantifies the angular acceptance.
As can be seen, more than half of the dark-matter particles are emitted toward the detector for small dark matter masses~$\mathcal{O}(10^{-2})~\mbox{GeV}-\mathcal{O}(10^{-1})~\mbox{GeV}$. However, this acceptance drops rapidly to the level of $\mathcal{O}(10^{-1})$ over the relevant mass range~$\mathcal{O}(10^{-1})~\mbox{GeV}-\mathcal{O}(1)~\mbox{GeV}$. The higher the production energy of the dark matter pair, the narrower the cone in which it propagates toward the detector. Consequently, the higher the beam energy, the better the detector acceptance.

\begin{figure}[tbh!]
	\center{\includegraphics[scale=0.475]{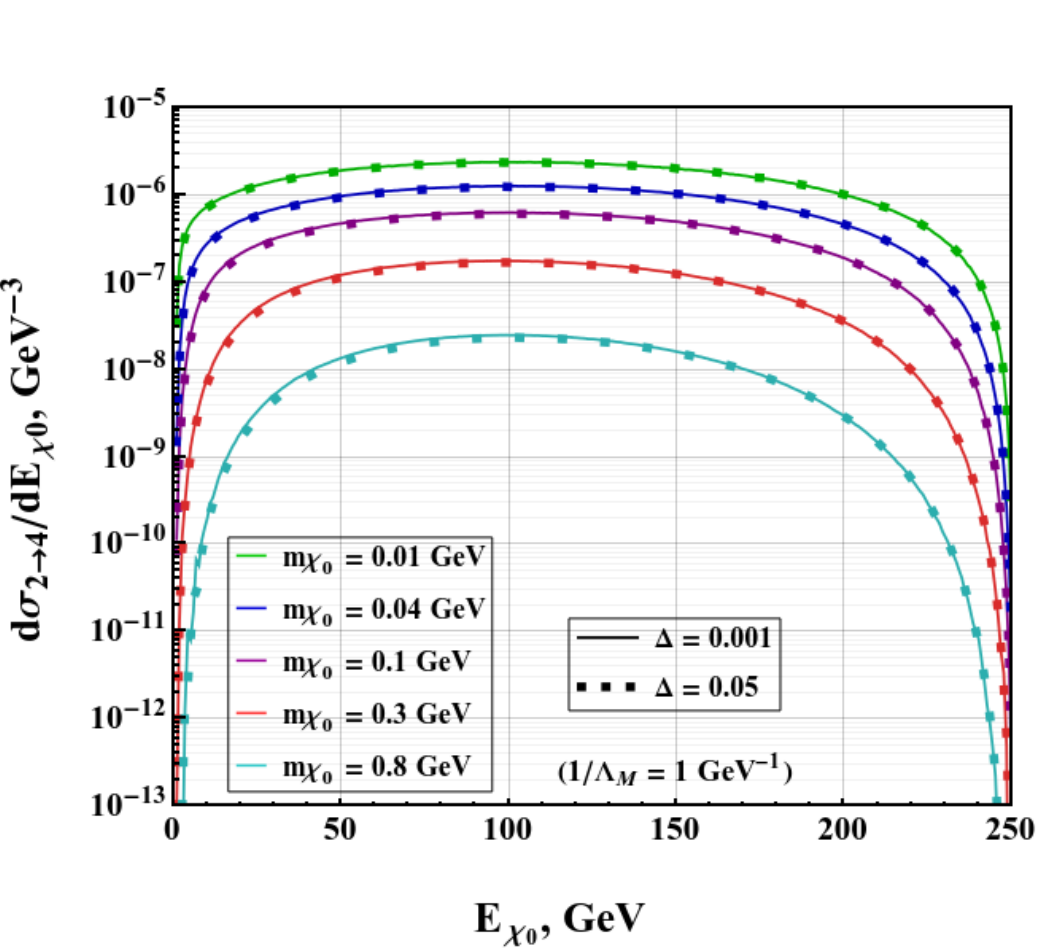}}
\caption{ 
Differential cross section of process~\eqref{eNtoeNChi0Chi1} as a function of the energy of particle~$\chi_0$, shown for different dark matter masses and mass splittings. Different colors indicate different dark matter masses, while solid and dashed curves correspond to mass splittings of 0.001 and 0.05, respectively.
}
	\label{fig:DiffCS_ICL_MiDM_Delta0x001}
\end{figure}

The differential cross section for the corresponding $2 \to 4$ process is shown in Fig.~\ref{fig:DiffCS_ICL_MiDM_Delta0x001} for various dark matter masses and mass splittings.
The mass splitting has only a mild impact on the energy distribution of dark-matter production. 
In particular, the largest variation in the differential cross section occurs for larger masses and is $<~\mathcal{O}(10)\%$.
For both primary beam energies, the total cross section ranges from~$\mathcal{O}(10^{5})~\mbox{pb}$ to~$\mathcal{O}(10^{2})~\mbox{pb}$ for the mass range~form 10~MeV to~1~GeV. 
The total cross section changes only at the $\mathcal{O}(1)\%$ level when the mass splitting is varied from~$\Delta~=~0.001$ to~$\Delta~=~0.05$.

\begin{table}[!th]
\begin{ruledtabular}
\begin{tabular}{cccc}
$m_{\chi_0}$, GeV & $\sigma_{2\to4}^{\rm cut}$, pb & $\sigma_{2\to4}^{\rm tot}$, pb & $\sigma_{2\to4}^{\rm cut}/\sigma_{2\to4}^{\rm tot}$ \\
$0.01$ & $1.10\times10^{5}$ & $1.77\times10^{5}$ & $0.62$ \\
$0.04$ & $4.64\times10^{4}$ & $1.04\times10^{5}$ & $0.44$ \\
$0.10$ & $1.82\times10^{4}$ & $5.97\times10^{4}$ & $0.31$ \\
$0.30$ & $3.00\times10^{3}$ & $2.12\times10^{4}$ & $0.14$ \\
$0.80$ & $2.01\times10^{2}$ & $3.99\times10^{3}$ & $0.05$ \\
\end{tabular}
\end{ruledtabular}
\caption{\label{tab:TCS2to4Ebeam125GeV}
Cross sections, $\sigma_{2\to4}^{\rm cut}$ and $\sigma_{2\to4}^{\rm tot}$, for the $2\to4$ process in the case of primary-beam energy $125~\mbox{GeV}$.}
\end{table}

\begin{table}[!th]
\begin{ruledtabular}
\begin{tabular}{cccc}
$m_{\chi_0}$, GeV & $\sigma_{2\to4}^{\rm cut}$, pb & $\sigma_{2\to4}^{\rm tot}$, pb & $\sigma_{2\to4}^{\rm cut}/\sigma_{2\to4}^{\rm tot}$ \\
$0.01$ & $1.45\times10^{5}$ & $2.12\times10^{5}$ & $0.68$ \\
$0.04$ & $7.29\times10^{4}$ & $1.35\times10^{5}$ & $0.54$ \\
$0.10$ & $3.44\times10^{4}$ & $8.26\times10^{4}$ & $0.42$ \\
$0.30$ & $8.69\times10^{3}$ & $3.41\times10^{4}$ & $0.25$ \\
$0.80$ & $1.08\times10^{3}$ & $9.00\times10^{3}$ & $0.12$ \\
\end{tabular}
\end{ruledtabular}
\caption{\label{tab:TCS2to4Ebeam250GeV}
Cross sections, $\sigma_{2\to4}^{\rm cut}$ and $\sigma_{2\to4}^{\rm tot}$, for the $2\to4$ process in the case of primary-beam energy $250~\mbox{GeV}$.}
\end{table}

We note that, the effective operator~\eqref{LagrangianMDM} provides a valid description of the interaction at energy scales below the electroweak symmetry-breaking scale ~\cite{Arina:2020mxo} (that implies the typical new physics scale $\Lambda_{\rm M} \simeq \mathcal{O}(10^2)~\mbox{GeV}$ ).
In particular, the momentum  vertex $\gamma^\ast~\to~\chi_0\chi_1$ is subject to the condition $Q_{2\to4}^2~<~m_Z^2$. For the process under consideration, the corresponding momentum satisfies:
$$
Q_{2\to4}^2 \le (Q_{2\to4}^{\max})^2 = \left(\sqrt{s_{2\to4}}-m_e-m_N\right)^2,
$$
where $s_{2\to4}~=~m_{e}^2~+~m_{N}^2~+~2m_{N}E_{\rm beam}$ in the lab frame. For an incident electron with energy $250~\mbox{GeV}$ scattering off an oxygen nucleus, the relation $Q_{2\to4}^{\max}~\simeq~72~\mbox{GeV}~<~m_{Z} \ll  \Lambda_{\rm M} \simeq \mathcal{O}(10^3)~\mbox{GeV}$ holds~\cite{Izaguirre:2015zva}.
Thus, for the experimental parameters under consideration, the effective operator remains valid~\cite{Jodlowski:2023ohn}. For a discussion of the ultraviolet-completed iDM scenario, see Ref.~\cite{Izaguirre:2015zva} and references therein.

\subsection{Scattering and decay of dark matter in ILC-BDX}

After the primary beam impinges on the target, a flux of boosted dark matter is produced.
Boosted dark matter can subsequently scatter off downstream electrons of the experimental facility as depicted in Fig.~\ref{FeynamProduction}~(b):
$$
 \chi_i(p) + e(k) \to   \chi_j(p') + e(k')
$$
where $q = k' - k$ and $Q^2 = -q^2$.
Assuming that the initial-state electron is at rest, the momentum transfer is given by $Q = \sqrt{2 m_e E_{\rm R}}~\ll~m_{Z}$, where $E_{\rm R} = E_e' - m_e$ is the electron recoil energy. 
For instance, retaining the leading terms in $Q, m_{\chi_0}, \Delta~\ll~E_{\chi_0}$, the differential cross section of the $\chi_0 e \to \chi_1 e$ process reads~\cite{Jodlowski:2023ohn}:
\begin{equation}\label{eq:dsdEchi0ScatDMandElectron}
    \frac{d\sigma_{\chi_0 e}}{dE_R}  
\simeq  
    \frac{4\alpha}{\Lambda_M^2 E_{\chi_0}^2}  
    \left[  
        \frac{E_{\chi_0}^2}{E_R}   
    +  
        \frac{m_{\chi_0}^2}{2m_e} 
    - 
        \Delta
        \frac{ m_{\chi_0}^2 E_{\chi_0}}{m_e E_{\rm R}} 
    \right].
\end{equation}

The mass splitting also makes the heavier dark matter state unstable, thereby opening two channels, $\chi_1 \to \chi_0 e^+ e^-$ and~$\chi_1~\to~\chi_0~\gamma$. 
The decay $\chi_1 \to \chi_0 e^+ e^-$ is suppressed by the phase space, and therefore two body monophoton decay (see Fig.~\ref{FeynamDecay}) implies $\mbox{Br}(\chi_1\to\chi_0\gamma)~\simeq~1$ for small relative mass splitting of light DM~\cite{Jodlowski:2023ohn}.
The decay length of $\chi_1$ in the lab-frame  can be expressed for sufficiently small splitting $\Delta \ll 1$, as~\cite{Gninenko:2026svu}: 
\begin{align}\label{DecLengthChi1}
 & d_{\chi_1} \simeq \;  1.91 \times 10^5 \; \text{m} 
\nn \\ &
\times \! \left(\! \frac{E_{\chi_1}}{10^2  \text{GeV}} \!\right)\! \!
\left(\! \frac{0.1 \text{GeV}}{m_{\chi_0}} \!\right)^{\!4}\!\!\! \left(\! \frac{0.001}{\Delta} \! \right)^3\!\!\! \left(\! \frac{\Lambda_{\rm M}}{10^3  \text{GeV}} \! \right)^2\!\!,
\end{align}


The primary dark-matter flux of $\chi_0$ leads to the scattering 
inside the detector with the typical probability:
\begin{equation}
    P_{\rm pr. sc.}^{\chi_0}
= 
    n_{\rm det}^e l_{\rm det} 
    \sigma^{\rm cut}_{2\to2} (E_{\chi_0}),
\end{equation}
where $\sigma^{\rm cut}_{2\to2}$ is obtained by integrating the differential cross section~\eqref{eq:dsdEchi0ScatDMandElectron} over recoil energies above the threshold $E_{\rm th}$.
The primary heavier dark-matter state can propagate to the detector and scatter with probability:
\begin{equation}
    P_{\rm pr. sc.}^{\chi_1}
\! = \!
    n_{\rm det}^e l_{\rm det} 
    \sigma^{\rm cut}_{2\to2}(E_{\chi_1})
    e^{-(L_{\rm sh} \! + \! L_{\rm dec.} \! + \! L_{\rm tg.})/d_{\chi_1}(E_{\chi_1})},
\end{equation}
where the exponential factor takes into account the 
fraction of survived $\chi_1$ that reaches the 
detector.
In particular, the dark matter flux traverses a $70$ m lead shielding volume on its way to the detector.  
For typical coupling $1/\Lambda_{\rm M}~=~10^{-3}~\mathrm{GeV}^{-1}$, the characteristic total cross section for dark matter scattering off electrons is given by~$\sigma_{\chi e}~\propto~4 \alpha/\Lambda_{\rm M}^2~\simeq~\mathcal{O}(10^{-35})~\mathrm{cm^2}$.
Accordingly, for the electron number density of lead, $n_e~\simeq~2.7~\times~10^{24}~\mathrm{cm^{-3}}$, the corresponding mean free path is of order~$\lambda_{\chi e}~=~(n_e\sigma_{\chi e})^{-1}~\simeq~\mathcal{O}(10^{8})~\mathrm{m}$.
Similarly, for scattering on nuclei we can write:
\begin{equation}\label{eq:lengthInShVol}
\lambda_{\chi N}~=~(n_N\sigma_{\chi N})^{-1}~\simeq~\mathcal{O}(10^{5})~\mathrm{m}.
\end{equation}
The flux of dark-matter particles reaching the detector can be reduced by less than $<\mathcal{O}(0.1)~\%$.
Therefore, the changes in the flux during dark matter passage through the shield can be safely neglected~\cite{Batell:2014mga}.

\begin{figure}[tbh!]
	\center{\includegraphics[scale=0.475]{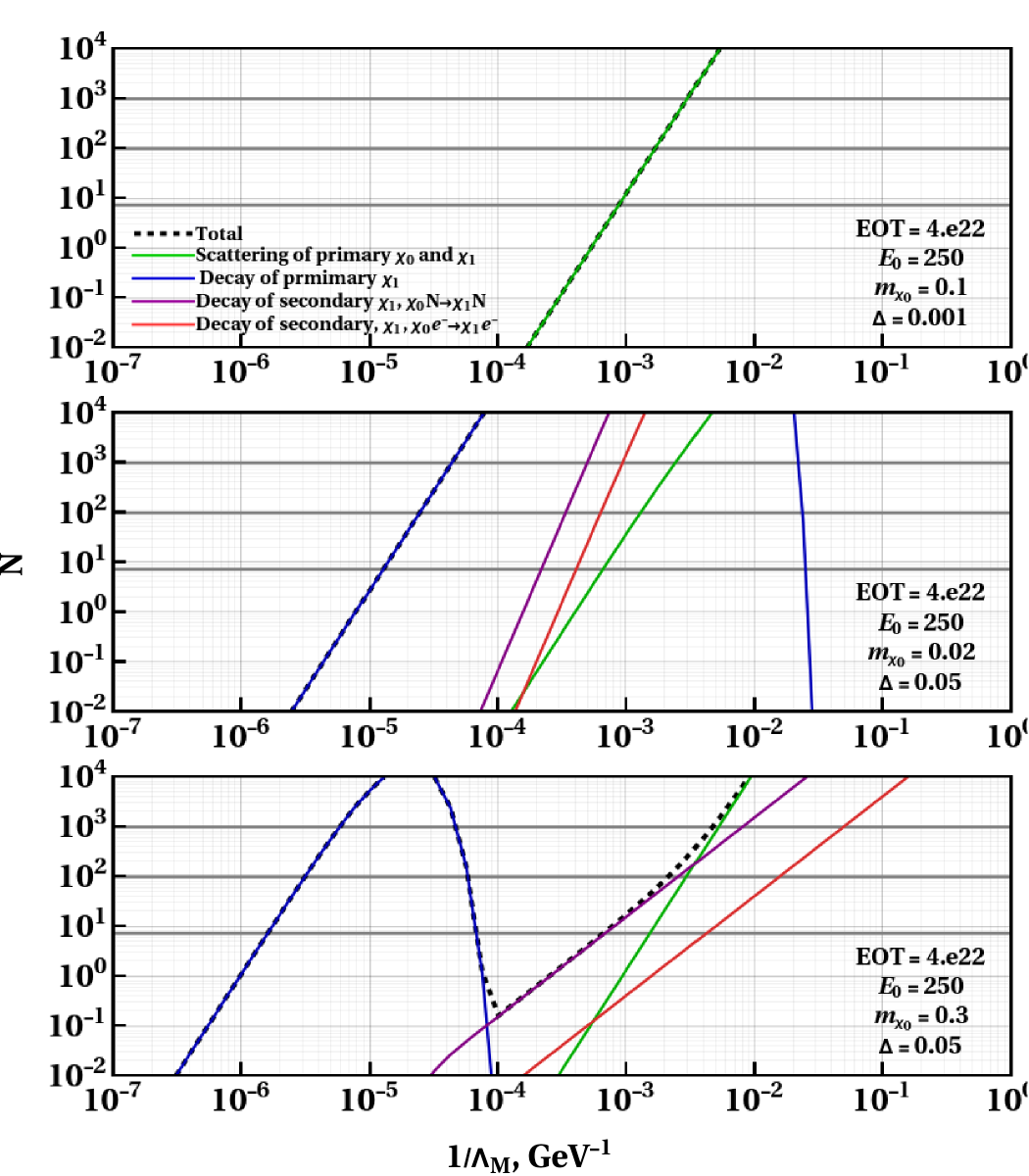}}
\caption{ 
Number of signal events for the different channels in the inelastic magnetic dipole dark matter model at the ILC-BDX experiment for the parameter values shown in the figure. 
In particular, the black dashed and solid gray curves correspond to the total number of events and the upper limit on the number of events, respectively. 
The solid green line corresponds to the direct detection, $\chi_i e \to \chi_j e$, of the primary produced iDM via $e N \to e N \gamma^*(\to \chi_1 \bar{\chi}_0)$.  The solid blue line corresponds to the detection of the  decays $\chi_1 \to  \chi_0 \gamma$  from the primary flux $e N \to e N \gamma^*(\to \chi_1 \bar{\chi}_0)$. The solid   violet (red) line corresponds the
 decay $\chi_1 \to  \chi_0 \gamma$ of the secondary heavy state produced via 
 up-scattering  
$\chi_0 N \to \chi_1 N$ ($\chi_0 e \to \chi_1 e$) in the shield material. 
}
	\label{fig:Nevent_ILCBDX}
\end{figure}

\begin{figure}[!tbh]
    \center{\includegraphics[scale=0.475]{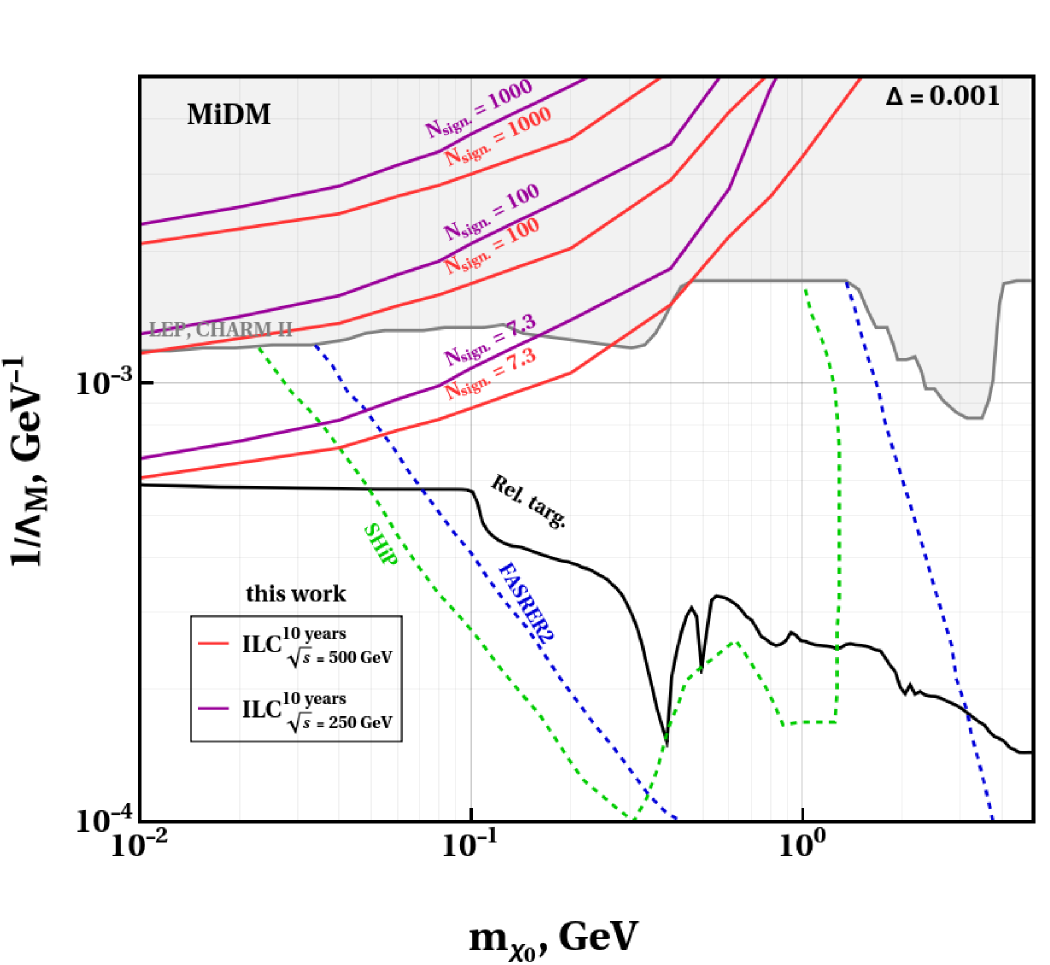}}
    \caption{ The expected limits at $95~\%$ CL on magnetic dipole dark matter coupling for the ILC-BDX fixed-target experiment as a function of the lightest DM state,~$\chi_0$ for mass splitting~$\Delta~=~0.001$.
The calculated projected constraints for the ILC-BDX experiment are shown as red and purple lines for primary beam energies of~$250~\mbox{GeV}$ and $125~\mbox{GeV}$, respectively, for 10 years of data taking.
We also show the current constraints from the LEP~\cite{L3:2003yon,Fortin:2011hv} and CHARM~II~\cite{CHARM:1985anb,CHARM:1983ayi} experiments as the gray shaded region~\cite{Kling:2022ykt}. 
The projected sensitivities of the SHiP and FASER2 experiments are shown by the blue and green dashed curves, respectively~\cite{Jodlowski:2023ohn}.
The relic-density target corresponding to the observed dark matter abundance is shown by the black curve~\cite{Jodlowski:2023ohn}.
}
    \label{fig:ICL_MiDM_compareDelta0x001}
\end{figure}

\begin{figure*}[!tbh]
    \center{\includegraphics[scale=1.0]{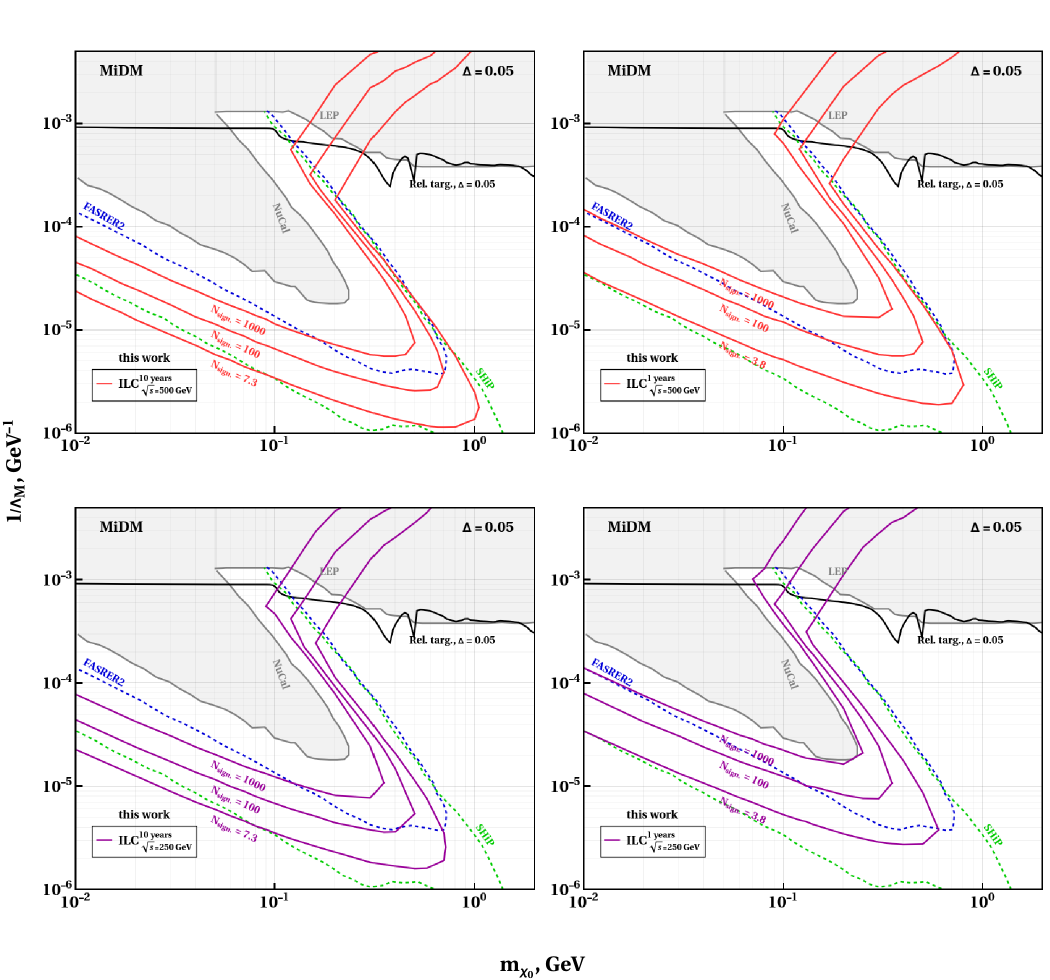}}
    \caption{ Same as fig.~\ref{fig:ICL_MiDM_compareDelta0x001}, but for~$\Delta~=~0.05$; the current constraints from the LEP and NuCal experiments are shown as the gray shaded region. 
}
    \label{fig:ICL_MiDM_compareDelta0x05}
\end{figure*}

The decay probability of primary $\chi_1$ within the fiducial volume of ILC-BDX with the energy cut $E_{\rm th}$ is~\cite{Jodlowski:2023ohn}:
\begin{align}\label{eq:ProbabilityDecayInDecayVol}
    & P_{\rm pr. dec.}^{\chi_1}(E_{\chi_1})  
= 
    A_E(E_{\chi_1})
\nn \\ & \times
    \left[
    e^{-(L_{\rm sh}\! + \! L_{\rm tg.})/d_i(E_i)}
    -
    e^{-(L_{\rm sh} \! + \! L_{\rm dec.} \! + \! L_{\rm tg.})/d_i(E_i)}
    \right],
\end{align}
where the energy threshold acceptance in the fiducial volume is defined as 
$$
A_E(E_{\chi_1}) = 	\frac{1}{\Gamma_{\chi_1}^{\rm tot}}
	\int d\Gamma_{\chi_1\to \gamma \chi_0}\,
	\Theta(E_\gamma^{\rm lab}-E_{\rm th})
$$


In addition to the decays of the $\chi_1$ states 
produced in the primary target, heavy states can also be generated via up-scattering of the lighter state in the shielding material~\cite{Jodlowski:2019ycu}.
The probability of decay after up-scattering within the decay volume is~\cite{Jodlowski:2023yne}:
\begin{align}
    &P_{\rm sec. dec.}^{\chi_1}(E_{\chi_1})
= 
	A_E(E_{\chi_1})
	\sum_{T=N,e}
	\frac{d(E_{\chi_1})}{L_{\rm int}^T(E_{\chi_1})}
\nn \\  \quad &  \times
	\left(
		1 - e^{-L_{\rm dec.}/d(E_{\chi_1})}
	\right)
	\left(
		1 - e^{-L_{\rm sh.}/d(E_{\chi_1})}
	\right).
\end{align}
where $L_{\rm int}^{T}(E_{\chi_1}) =	1/(n_T \sigma_T(E_{\chi_1}))$ is interaction length of the scattering channels $T=(N,e)$.
The total cross section for dark matter scattering off nuclei can be written as~\cite{Jodlowski:2023ohn}:
$$
\sigma_N(E_{\chi_1})
\simeq
\frac{4\alpha_{\rm EM}Z^2}{\Lambda_M^2}
\left[
\log
\left(
\frac{d}{a^{-2}-t_{\rm max}}
\right)
-2
\right],
$$
where we introduce
$$
	a = \frac{111}{m_e Z^{1/3}},
\quad
	d = \frac{0.164}{A^{2/3}}{\rm GeV}^2,
\quad
	t_{\rm max}
	\simeq
	-\frac{m_0^4+m_1^4}{4E_{\chi_1}^2}.
$$
The corresponding expression for the differential cross section for scattering off electrons is given in Eq.~\eqref{eq:dsdEchi0ScatDMandElectron}.
Decays of $\chi_1$ states produced via $\chi_0$ up-scattering can yield stronger constraints than decays of heavy states from the primary flux in the case of short-lived $\chi_1$, owing to their production closer to the detector~\cite{Jodlowski:2019ycu}.

For secondary production in the shielding volume, we employ the collinear approximation, in 
which particles propagating within the detector angular cone before scattering are assumed to 
remain directed toward the detector after scattering. 
A full treatment of the angular distribution in secondary scattering, including the 
corresponding geometric acceptance of the detector, requires dedicated Monte-Carlo 
simulations and is beyond the scope of the present work.

%

Thus, the number of signal events for inelastic magnetic dipole dark matter in the ILC-BDX experiment is given by
\begin{align}\label{eq:NumberOfSignalEvents}
   & N_{\rm sign.} 
\simeq 
    \mbox{EOT} \times
    \frac{\rho N_A}{A} \times L_T 
    \int\limits_{E_{\rm min}}^{E_{\rm max}} dE_{\chi_1}
    \frac{d\sigma_{2\to 4}^{\rm cut}}{dE_{\chi_1} }
\nn \\ &    
 \times   \left( 
        P_{\rm pr. sc.}^{\chi_1}
    +   P_{\rm pr. sc.}^{\chi_0}
    +   P_{\rm pr. dec.}^{\chi_1}
    +   P_{\rm sec. dec.}^{\chi_1}
    \right) 
    ,
\end{align}
where we assume that the spectra of the mass eigenstates are identical.
Indeed, in high-energy fixed-target experiments, the effect of the mass splitting on the spectrum is small.
We also neglect the probability of dark matter scattering in the shielding material~(see~Eq.~\eqref{eq:lengthInShVol}) when evaluating signal events induced by scattering in the detector material.

In Fig.~\ref{fig:Nevent_ILCBDX}, we show the number of signal events for the different channels at the ILC-BDX experiment for the planned statistics~$4\times10^{22}$ and a primary beam energy of 250 GeV.

In upper panel of Fig.~\ref{fig:Nevent_ILCBDX}, we set a mass splitting of~$0.001$ and a dark-matter mass of 100 MeV, for which the dominant contribution arises from scattering of the primary dark-matter flux in the detector.  For this benchmark, the decay channels do not lead significant yield over the entire mass range from 10 MeV to 1 GeV due to the energy cut, $E_{\rm th}=1~\mbox{GeV}$.
Moreover, for small mass splitting, the decay length of the heavier state is much larger than the detector size, that can lead to signal events of the $\chi_1$ scattering in detector.

The middle and bottom panel of 
Fig.~\ref{fig:Nevent_ILCBDX} correspond to a mass splitting of $0.05$. 
In the second row, for a mass of 20 MeV, the strongest constraints are provided by the decay channel of the heavier state from the primary flux. 
Although for sufficiently large coupling values,~$1/\Lambda_{\rm M}~\simeq~\mathcal{O}(10^{-2})$, the heavier state becomes short-lived and no longer reaches the decay volume. The combination with channels involving secondary production of~$\chi_1$ followed by its decay excludes coupling values in this region. 
As a result, a one-sided constraint on the coupling emerges~$1/\Lambda_{\rm M}\lesssim \mathcal{O}(10^{-5})$.
One can also see that up-scattering of the primary dark matter flux on electrons in the shielding volume, followed by decay, leads to weaker constraints than the analogous process on nuclei because of the $\sigma_{\chi N} \propto Z^2$ enhancement.

In bottom panel of Fig.~\ref{fig:Nevent_ILCBDX}, for a mass of 300 MeV, the constraints on the coupling arise from decays in the small-coupling region as~$1/\Lambda_{\rm M} \lesssim \mathcal{O}(10^{-6})$ and $1/\Lambda_{\rm M}\gtrsim \mathcal{O}(10^{-5})$. Whereas for sufficiently large couplings~$1/\Lambda_{\rm M}\lesssim \mathcal{O}(10^{-3})$ the dominant constraints are provided by decays of secondarily produced $\chi_1$ states. 
In this case, however, an open parameter region remains between these two regimes. 
Indeed, in this region the particles are short-lived enough not to traverse the shielding volume, but still long-lived enough that dark matter particles produced inside the shielding volume decay outside the detector. 
For these parameters, the constraints on the coupling take the form of an excluded interval together with an upper bound.

A detailed analysis of the additional 
 signatures~\cite{Schuster:2021mlr,Asai:2021ehn,Asai:2023dzs,Essig:2024dpa} 
 of iDM  at ILC-BDX we leave to future work.

\section{The experimental reach \label{sec:ExpectedReach}}


In this section, we discuss the sensitivity of the ILC-BDX fixed-target experiment to the thermal inelastic magnetic dipole dark matter model in the mass range \mbox{$10~\mbox{MeV} \lesssim m_{\chi_0} \lesssim 1~\mbox{GeV}$}. 

The corresponding experimental reach are shown in the Fig.~\ref{fig:ICL_MiDM_compareDelta0x001} for mass splittings as~\mbox{$\Delta = 0.001$} and in the Fig.~\ref{fig:ICL_MiDM_compareDelta0x05} for~\mbox{$\Delta = 0.05$}.
We focus on small mass splittings as~\mbox{$ \mathcal{O}(10^{-3}) \lesssim \Delta  \lesssim \mathcal{O}(10^{-2})$}, which correspond to the viable region of parameter space of the considered model~\eqref{LagrangianMDM}.
Based on the analytical approximation, we displays the standard thermal target lines for two representative mass splittings, $\Delta = 0.001$ and $\Delta = 0.05$. 


To derive the constraints, we use the~$95\%$~CL upper limits,~\mbox{$\Npup = 3.8$} and~\mbox{$\Npup = 7.3$}, for the 1-year and 10-year of data taking, respectively, as discussed in Sec.~\ref{sec:ExperimentSetup}. 
We also account for uncertainties in the  background and in the final detector design by considering benchmark signal yields of 100 and 1000 events.
To calculate the number of signal events, we use Eq.~\eqref{eq:NumberOfSignalEvents}, and take into account: (i) interactions of the primary dark matter flux in the detector volume, (ii) decays of the heavy state from the primary flux, (iii) decays of the heavy state produced via up-scattering on nuclei, (iv) decays of the heavy state produced via up-scattering on electrons.
For secondary production of the heavy dark matter state, scattering off nuclei in the shielding volume provides the dominant contribution because of the $Z^2$-enhancement.


We consider primary beam energies of~\mbox{$E_{\rm beam} = 250~\mbox{GeV}$} and~\mbox{$E_{\rm beam} = 125~\mbox{GeV}$} as benchmark values. 
We find that the bound on the interaction coupling becomes stronger by factor of $\mathcal{O}(1)$ for masses in the range from~$10~\mbox{MeV}$ to~$1~\mbox{GeV}$ when the primary beam energy is doubled.
At small masses, the weaker constraints for a primary beam energy of~\mbox{$E_{\rm beam} = 125~\mbox{GeV}$} arise from the logarithmic energy dependence of the total cross section. 
At larger masses, lowering the primary beam energy from~\mbox{$E_{\rm beam} = 250~\mbox{GeV}$} to~\mbox{$E_{\rm beam} = 125~\mbox{GeV}$} likewise weakens the constraints because of the reduced acceptance, as seen in Tabs.~\ref{tab:TCS2to4Ebeam125GeV} and~\ref{tab:TCS2to4Ebeam250GeV}.


For a mass splitting of $\Delta = 0.001$, we show only the projected sensitivity, as presented in Fig.~\ref{fig:ICL_MiDM_compareDelta0x001}.  
The heavy dark matter state, $\chi_1$, produced in the target can propagate to the detector and  create a signal event.
The probability of $\chi_{1}$ reaching the detector is characterized by an exponential factor with decay length~\eqref{DecLengthChi1}.
In particular, for $\Delta = 0.001$ the dominant portion of  $\chi_1$ can reach the detector for the mass range \mbox{$10~\mbox{MeV} \lesssim m_{\chi_0} \lesssim 1~\mbox{GeV}$}, thus  $P_{\rm det.}^{\chi_1} \simeq  P_{\rm det.}^{\chi_0}$. 
For this mass splitting, the energy of the photon emitted in the decay of the heavier state 
is below the detector energy threshold imposed for background rejection. 
Consequently, for this mass splitting, the dominant channel is scattering of dark matter 
particles from the primary flux off atomic electrons. 

In particular, the lower photon-energy thresholds of experiments FASER2, and SHiP lead to that this experiments are sensitive to decays of the heavier state for mass splitting $\Delta = 0.001$.
The resulting constraints depend only weakly on the mass splitting for $\Delta < 0.001$ due to the sufficiently high primary-beam energy and the decay length as $d_{\chi_1}~>~\mathcal{O}(10^{5})~\mbox{m}$. 
Increasing the required number of signal events by one order of magnitude weakens the bound on the interaction coupling by a factor of 1.8. 

Thus, assuming optimistic case of expected signal events $ N_{\rm sign }\simeq \mathcal{O}(10)$ and the 10-years statistics, the ILC-BDX can set new bounds on the interaction coupling at the level of $1/\Lambda_{\rm M} \lesssim 5\cdot10^{-4}~\mbox{GeV}^{-1}-1.1\cdot10^{-3}~\mbox{GeV}^{-1}$ in the mass range from 10 MeV to 100 MeV.


We now discuss the constraints for a mass splitting of $\Delta = 0.05$, shown in Fig.~\ref{fig:ICL_MiDM_compareDelta0x05}. 
The lower branch of bound associated with decays of the heavy state from the primary flux scales as $N_{\rm sign.}\propto (1/\Lambda_{\rm M})^4$, since the heavy state is sufficiently long-lived.
In this regime, the lower contour of the bound on the interaction coupling scales as $1/\Lambda_{\rm M}\propto (N_{\rm sign.})^{1/4}$.
By contrast, the upper branch of the constraints from decays of the heavy state in the primary flux depends  weakly on the experimental parameters due to to the short lifetime of the $\chi_1$ state. 
Indeed, in this case the differential number of events is exponentially suppressed by the ratio of the shielding length to the decay length of the heavy state. 
Therefore, increasing the number of signal events or statistics by one order of magnitude weakens the lower  branch of the constraint in the small-coupling region by a factor of 1.8.

For $\Delta = 0.05$, the existing bounds exclude the iDM thermal target curve for masses below 50 MeV, based on the results of the NuCal experiment. 
The ILC-BDX experiment can extend this reach and exclude the parameter scape of iDM up to masses of about $300~\mbox{MeV}$  in  with a 10-year statistics and a primary beam energy of 250 GeV with optimistic case of expected number of signal events $ N_{\rm sign }\simeq \mathcal{O}(10)$.
Similarly, for reduced beam energy or smaller statistics, the corresponding reach in mass decreases to about $200~\mbox{MeV}$.

In the  mass range, from $\mathcal{O}(100)~\mbox{MeV}$ to $\mathcal{O}(300)~\mbox{MeV}$, the constraints on thermal iDM arise from decays of the heavy state in the primary flux and from decays of the heavy state produced inside the shielding volume.

The ILC-BDX bounds derived from decays of the heavy state in the primary 
flux are comparable to that of the NuCal, FASER2, and SHiP experiments. 
In particular, for a beam energy of 250 GeV, maximal statistics, the resulting bounds are comparable to those expected from SHiP, assuming $N_{\rm sign. }\simeq 7.3$.
On the other hand, in the most conservative case, corresponding to a beam energy of 125 GeV, a 1-year exposure, and an expected signal yield of 1000 events, the resulting bounds are at the level of NuCal.
Thus, for $\Delta = 0.05$, the ILC-BDX experiment can probe the mass range from 10 MeV to 1 GeV for interaction couplings in the range $10^{-6} ~\mbox{GeV}^{-1}\lesssim 1/\Lambda_{\rm M} \lesssim 10^{-3}~\mbox{GeV}^{-1}$, depending on the experimental setup.


\section{Conclusion\label{sec:Conclusion}}

In our analysis, we show that the ILC-BDX experiment is sensitive to thermal inelastic 
magnetic dipole dark matter with sub-GeV masses through dark matter scattering off 
detector electrons and the decay of heavy state of dark matter.
In particular, for $\Delta = 0.05$, the experiment yields constraints comparable to those from NuCal, FASER2, and SHiP.
We conclude that the ILC-BDX experiment can potentially probe the parameter space of this 
model, especially for  mass splittings of $\Delta \gtrsim 0.05$ and  light DM, 
$m_{\chi_0} \lesssim \mathcal{O}(300)~\mbox{MeV}$.
For $\Delta = 0.001$, the experiment can set new bounds in the mass range from 10 MeV to 100 MeV for the coupling at the level of $1/\Lambda_{\rm M} \lesssim 5\cdot10^{-4}~\mbox{GeV}^{-1}-1.1\cdot10^{-3}~\mbox{GeV}^{-1}$.

\begin{acknowledgments} 
The work of IV and DK  was supported by the Foundation for the Advancement of Theoretical Physics and Mathematics BASIS (Project No.~\text{24-1-2-11-2} and No.~\text{24-1-2-11-1}).
This work was supported by The Ministry of Science and Higher Education of the Russian Federation in part of the Science program (Project No. FSWW-2026-0046).
\end{acknowledgments}

\appendix

\bibliography{bibl}

\end{document}